\documentclass[
aps,%
final,%
notitlepage,%
oneside,%
onecolumn,%
nobibnotes,%
nofootinbib,
superscriptaddress,%
showpacs%
]%
{revtex4}

\usepackage{graphicx}%
\usepackage{dcolumn}
\usepackage{amsmath}

\begin{document}
\title{Lattice Relaxation and Charge-Transfer Optical Transitions
Due to Self-Trapped Holes in Non-Stoichiometric LaMnO$_3$ Crystal}

\author{N. N. Kovaleva}
\email{nkovalev@issp.ac.ru}
\affiliation{Institute of Solid State Physics, Russian Academy of Sciences, 
Chernogolovka, Moscow distr., 142432 RUSSIA}  
\affiliation{University College London, Gower Street, London WC1E 6BT, UK}  

\author{J. L. Gavartin}
\author{A. L. Shluger}
\affiliation{University College London, Gower Street, London WC1E 6BT, UK}  

\author{A. V. Boris}
\affiliation{Institute of Solid State Physics, Russian Academy of Sciences, 
Chernogolovka,
Moscow distr., 142432 RUSSIA}  
\affiliation{Max-Planck-Institut f\"{u}r Festk\"{o}rperforschung, 
Heisenbergstr. 1, 70569 Stuttgart, Germany}  

\author{A. M. Stoneham}           
\affiliation{University College London, Gower Street, London WC1E 6BT, UK}  

\date{\today}

\begin{abstract}  
We explore the role of electronic and ionic polarisation energies in the 
physics of the "colossal" magnetoresistive (CMR) materials. We use the 
Mott-Littleton approach to evaluate polarisation 
energies in LaMnO$_3$ lattice associated with holes localized on both 
Mn$^{3+}$ cation and O$^{2-}$ anion. The full (electronic and ionic) lattice 
relaxation energy for a hole localized at the O-site is estimated as 
2.4 eV which is appreciably greater than that of 0.8 eV for a hole localized 
at the Mn-site, indicating on the strong electron-phonon interaction in
the former case. The ionic relaxation around the localized holes differs for 
anion and cation holes. That associated with Mn$^{4+}$ is approximately 
isotropic, whereas ionic displacements around O$^-$ holes show axial symmetry, 
the axis being directed towards the apical oxygens.
\par
Using a Born-Haber cycle we examine thermal and optical energies of the
hole formation associated with electron ionization from 
Mn$^{3+}$, O$^{2-}$ and La$^{3+}$ ions in LaMnO$_3$ lattice. 
For these calculations we derive a phenomenological value for 
the second electron affinity of oxygen in LaMnO$_3$ lattice 
by matching the optical energies of La$^{4+}$ and O$^-$ hole 
formation with maxima of binding energies in 
the experimental photoemission spectra. 
The calculated thermal energies predict that the electronic hole is 
marginally more stable in the Mn$^{4+}$ state in LaMnO$_3$ host lattice, 
but the energy of a hole in the O$^-$ state is only higher by a small amount, 
0.75 eV, rather suggesting that both possibilities should be treated seriously. 
\par
We examine the energies of a number of fundamental optical transitions,
as well as those involving self-trapped holes of Mn$^{4+}$ and O$^-$ 
in LaMnO$_3$ lattice. The reasonable agreement with 
experiment of our predicted energies, linewidths and oscillator strengths 
leads us to plausible assignments of the optical bands observed. We deduce 
that the optical band near 5 eV is associated with O(2p) - Mn(3d) transition
of charge-transfer character, whereas the band near 2.3 eV is rather 
associated with the presence of Mn$^{4+}$ and/or O$^-$ self-trapped holes in 
non-stoichiometric LaMnO$_3$ compound. 

\end{abstract}

 

\maketitle   
 
\section{Introduction}    
The striking behaviour of the CMR oxides of $\rm R_{1-x}A_xMnO_3$ 
(R: trivalent rare-earth ions, A: divalent alkaline-earth ions, 
$0.2 \leq x \leq 0.5$) arises from the interplay of several  
distinct energy terms: magnetic interactions, electronic band 
structure energies, crystal field splittings, vibrational energies 
and electron lattice coupling, including small polaron ideas and 
the Jahn-Teller (JT) effect.
Understanding this behaviour has been helped very greatly by the 
use of models to map the various regimes of behaviour \cite{Millis}. 
The experimental evidence \cite{Ju} suggests that manganites are doped
charge-transfer insulators having O(2p) holes as the current carriers rather
than Mn$^{3+}$ (3d) electrons. However, whether holes reside on O- and/or
Mn- sites is still the subject of controversy.  
Some of the models of polarisation and vibration in CMR systems 
make major approximations, such as a single vibrational frequency 
(Einstein model) or rigid, unpolarizable ions. These simplifications 
are known to give seriously inadequate results, both quantitatively 
and qualitatively. For example, for the charge-transfer (CT) transitions 
of the zinc vacancy centre V$^-$ in ZnSe, optical spectroscopy 
\cite{ZnSe opt spectr} allows one to obtain the key relaxation and 
tunnelling energies. But these values, in the simple one frequency, 
rigid-ion, model are inconsistent with the observed localization 
of the charge on a single Se neighbour to the vacancy 
\cite{Harding & Stoneham 1982}. However, if one goes to a general 
model at the harmonic and dipole approximation level, namely the 
shell model, there is both consistency and good agreement with 
experiment. What the shell model 
\cite{Dick & Overhauser 1958,Cochran 1971} does which is significant 
is, first, to separate the ionic and electronic polarisations 
properly, so that phonons are predicted well and polarisation at 
the atomic scale is well reproduced. Secondly, the shell model 
recognizes that the local environment affects the polarisability 
of ions through short-range repulsive forces. As a result, the 
shell model provides a strong framework for understanding energies, 
which are dominated by polarisation and distortion. Such energies 
include those describing small polarons [3-5] 
\nocite{Norgett & Stoneham 1973,Stoneham 1989,
Shluger & Stoneham 1993} and optical CT transitions (as considered 
for MgO \cite{Stoneham & Sangster(1980)} and V$^-$ centres in ZnSe 
\cite{Harding & Stoneham 1982}). The shell model has 
also been used extensively in studies of defect energetics and 
non-stoichiometry in oxides \cite{Catlow}. Its considerable 
quantitative success arises largely because it provides such an 
accurate description of the large polarisation energies.
\par
It is helpful to recognize the orders of magnitude of the several 
energy terms for CMR oxides. Obviously, a small energy does not 
mean that the particular energy is unimportant, but a small 
value often means that very simple ideas for those terms are 
sufficient when examining phenomena dominated by large energies. 
Typical magnitudes are these:   
\begin{tabbing}   
Crystal field splitting energies (from data on many systems 
\hspace{2.5cm} \= 1 eV typical  \kill    
CMR instability energy of an electron in an external field of 10 T \> 0.001 eV 
($ \sim \mu gH$), \\   
magnetic exchange (from $kT_N$, $T_N$ being the Neel temperature) \>   0.01   
eV, \\ 
energy of non-cubic structural deformation of LaMnO$_3$ cell\> $\leq$ 0.4   eV, 
\\
Jahn-Teller energy (from largest known JT energies) \> $\leq$ 0.4  eV, \\ 
crystal field splitting energies (from data on many systems) \> 1  eV typical,\\
polarisation energies (net charge $\pm e$) \>  5 to 10 eV, \\ 
free ion ionisation potentials   \> Tens of eV, \\ 
Madelung energies (fully ionic models)  \>  Tens of  eV.   
\end{tabbing} 
\par
In this paper we shall be concerned mainly with the polarisation 
energies, for which the large energy terms are dominant. 
We shall only discuss Jahn-Teller and crystal field (CF) 
energies in simple terms, although we remark that one-frequency 
models of the JT effect will also lead to inconsistencies.  
\par
We apply the shell model calculations to look specifically 
at energies associated with the localized holes of Mn$^{4+}$ 
and O$^-$ in non-stoichiometric or slightly doped "parent" 
LaMnO$_3$ compound. Using this model, we address some of the 
issues in physics of CMR systems for which the polarisation 
energies are crucial. First, we calculate the electronic and 
ionic polarisation energies due to holes localized on Mn$^{3+}$  
and O$^{2-}$ ions in order to estimate the key polaron energies and 
examine the controversial question as to whether holes reside 
at Mn- or O-sites in LaMnO$_3$ lattice. Second, we estimate the 
energies of the main CT transitions including  Mn$^{4+}$ and  O$^-$ 
species, which determine specific transport properties of doped CMR materials.
We analyze their contribution to the optical conductivity in the 
non-stoichiometric LaMnO$_3$ crystal and make the assignment of the bands
in the optical conductivity spectrum more clear-cut.
 
\section{Description of $\rm LaMnO_3$ system and shell model approximation}

Many of the CMR materials are hole-doped systems of perovskite 
manganites of the form $\rm La_{1-x}A_xMnO_3$. Their properties 
are  intimately related to those of the "parent" compound (x=0). 
LaMnO$_3$ is an A-type antiferromagnet below the $T_N \simeq $ 140 K, 
in which the MnO$_2$ ferromagnetic layers are stacked along the 
c-axis with alternating spin directions. The structure of the 
perovskite manganites can be clearly understood starting from the 
simple cubic perovskite structure ({\it Pm3m}). The idealized cubic structure 
of LaMnO$_3$ featuring a chain of the corner-sharing MnO$_6$ 
octahedra is presented in Figure 1. The Mn$^{3+}$ ion with the 
3d$^4$ electronic configuration is known to exhibit a large 
Jahn-Teller effect in other systems \cite{Englman}. 
Therefore it is natural to assume that the JT instability of the 
Mn$^{3+}$ ion can contribute to an orthorhombic distortion of the 
perovskite structure of $Pnma$ symmetry in LaMnO$_3$ crystal. 
The orthorhombic structure can be obtained from the cubic 
perovskite structure by the two consequent and coordinated rotations 
of the MnO$_6$ octahedra around the [010] and [101] directions, 
as shown in Figure 1. Another possible contribution to the observed 
distortion from cubic symmetry in LaMnO$_3$ could be attributed 
to an atomic size mismatch: the sum of the Mn-O layer ionic radii, 
$r_{Mn} + r_O$, does not match that of the La-O layer, 
$(r_{La }+ r_O)/ \sqrt 2 $, in the right way for a stable cubic 
structure. Size mismatch effect is known to be a common reason 
for distortions in different perovskite oxides. 
Our shell model calculations performed for LaMnO$_3$ {\it Pnma} 
structure point out that the orthorhombic distortions experimentally 
observed at low temperatures could not be caused by simply lattice mismatch 
effect (which, in principle, must be properly described in the framework of 
the shell model approximation), but rather caused by the both effects, 
with a comparative contribution from the JT effect. Some special efforts 
should be undertaken to account for the JT effect empirically 
in the framework of the shell model. We perform the shell model calculations 
for the cubic perovskite structure (Figure 1) in the present work. 
This approximation seems to be mostly relevant 
to the non-magnetic quasicubic perovskite structure of LaMnO$_3$ crystal
experimentally observed at high temperatures T $\geq$ 400 K $>$ T$_N$ $\simeq$ 140 K. 
We suggest our modeling of the cubic perovskite structure provides 
a reasonable model as we are mainly interested to estimate the key polarization
energies associated with polaron-type charge carriers in the high-temperature insulating 
quasicubic phase of the CMR lattices.
\par
We model the LaMnO$_3$ system using methods based on the shell model  
and Mott-Littleton approach which have successfully 
been applied to study properties of a wide range of oxides 
(including transition metal oxides), halides and other systems 
\cite{Shluger & Stoneham 1993, Catlow2}. 
The calculations are performed using the GULP code \cite{Gale}. 
In the shell model  \cite{Dick & Overhauser 1958}
the lattice is considered as an assembly of
polarizable ions, represented by massive point cores 
and massless shells coupled by isotropic harmonic forces. 
Interacting potential includes contributions from 
Coulomb, polarization and short-range interactions.  
We adopt a fully ionic model (with formal charges of ions 
in the LaMnO$_3$ lattice: La$^{3+}$, Mn$^{3+}$, and O$^{2-}$). 
This is less restrictive than one might think, 
since a parallel covalent description is possible \cite{Catlow & Stoneham 1982}.
The sum of the core 
and shell charges is equal to the formal charge of the ion in 
the lattice. The core and shell charges and the spring constant 
of each ion are parameters of the model. 
The electronic polarisation of the ions is represented 
by the displacement of their shells relative to the cores in the 
dipole approximation.
The lattice distortion is simulated by the core displacements 
from their lattice site positions. 
\par
In our model cations are treated unpolarizable and short-range 
interactions between the relatively small cations (core-core) 
are ignored. The short-range potentials used for the 
shell-shell (oxygen-oxygen) and core-shell (metal-oxygen) 
interactions are of the Buckingham form:  

\begin{eqnarray}
  V_{ij} = A_{ij} \exp (- r / \rho_{ij} ) - C_{ij} / r^6   
  \label{eq1}
\end{eqnarray}

The parameters of both repulsive and attractive components of 
Buckingham potential for shell-shell (O$^{2-}$ : O$^{2-}$) interactions 
used in this work are obtained in ref. \cite{Stoneham} and presented 
in Table 1, (a). 
The Buckingham parameters for the core-shell Mn$^{3+}$ : O$^{2-}$ and 
La$^{3+}$ : O$^{2-}$ interactions 
were fitted in this work using the experimental data including 
the lattice parameter, the static and high-frequency dielectric 
constants, and the 
frequencies of the transverse optical phonons in LaMnO$_3$ crystal
\cite{Arima and Tokura}. The dielectric constants 
are especially important if one wishes to predict polarisation 
energies accurately. We have not found out an experimental value 
of the static dielectric constant of LaMnO$_3$ in the literature.  
We are grateful to T. Arima and Y. Tokura \cite{Private} 
for sending us the experimental data on the reflectivity spectra of 
LaMnO$_3$ measured at room temperature and reported in ref. 
\cite{Arima and Tokura}. The experimental value of the static dielectric
constant $\epsilon_0 \approx 18 \pm 2$ was derived from these data 
in the present work by the Kramers-Kronig analysis, 
and further used in the fitting procedure. 
The parameters fitted for LaMnO$_3$ ({\it Pm3m}) in ref. \cite{Islam} 
(see Table I, (b)) were used as the starting values 
for the core-shell La$^{3+}$ : O$^{2-}$ and Mn$^{3+}$ : O$^{2-}$ 
short-range interaction potentials. 
The oxygen shell charge was taken of -2.48 $\mid e \mid$ and the 
shell-core spring constant $k$ was chosen to give the correct 
value for the static dielectric constant $\epsilon_0$. 
\par 
The final values of our shell-model parameters are 
presented in Table I, (a). 
The calculated and experimental properties of LaMnO$_3$ 
({\it Pm3m}) are summarized in Table II. One can see that both 
sets of parameters (Table I, (a,b)) give close values for 
the lattice parameter and cohesive energy, however, our parameters 
at the same time give results close to the static and 
high-frequency dielectric constants. 
The value of the static dielectric constant calculated with the parameters 
of ref. \cite{Islam} is much higher than that derived from the 
experimental reflectivity spectra. Our model also agrees well 
with the experimental values of the transverse optical phonon 
energies \cite{Arima and Tokura}. The phonon bands obtained in 
our calculations correlate well with those observed with higher 
oscillator strengths. In particular, the predicted phonon energies 
agree well for the La-external ($\omega_{TO_1}$), Mn-O-Mn bending mode 
($\omega_{TO_2}$) and Mn-O stretching mode ($\omega_{TO_3}$) for 
the quasicubic perovskite structure of the strongly doped perovskite 
manganite system $\rm La_{0.67}Ca_{0.33}MnO_3$ \cite{Boris1,Boris2}. 
\par
We have also tested yet another set of short-range pair potentials 
which are different for Mn ion in different valence states 
Mn$^{2+}$, Mn$^{3+}$, and Mn$^{4+}$. 
They were obtained by fitting the equilibrium structures of 
several oxide compounds, such as MnO, LaMnO$_3$, and Ca$_2$MnO$_4$ 
\cite{Robin}. We tested pair-potentials for Mn$^{4+}$ and 
Mn$^{3+}$ from this set, presented in Table I, (c). These parameters 
also give good results (see the set of values (c) in Table II) for 
the lattice parameter and the dielectric constants, but are less 
successful in predicting the optical phonon frequencies. As will 
be shown below, both these and our parameters give similar values 
for the calculated properties of polarons in these crystals 
validating the correctness of the shell model approach.
\par
We apply then the shell model parameters to estimate key defect 
energies using the well-known Mott-Littleton method (see ref. \cite{Catlow2} 
for more detailed description). It is based on the concept that the total 
energy of the crystal lattice containing defect is minimized by a relaxation 
of the ions surrounding the defect, and this relaxation decreases fairly 
rapidly for distances away from the defect. In these calculations, the crystal 
is divided into three regions: an inner spherical region I, containing 
the defect and its immediate surroundings, an intermediate finite region II, 
which is created to link properly region I and an outer infinite region III, 
which responds as a dielectric continuum. The finite regions I and II are 
embedded in the infinite region III. The typical radii of regions I and II 
used in our calculations were 10 and 25 \AA, respectively. 
We considered an electronic hole located in the centre of region I, 
which is the most perturbed. The displacements of cores and shells in this 
region are calculated explicitly. In intermediate region II the ions are also 
treated in the shell model, but their displacements and polarisations 
are derived from the dielectric continuum approximation. 
The system total energy is minimized (the preset accuracy was of 0.01 eV) 
with respect to the positions of all cores and shells in regions I and II 
in the potential produced by the polarized region III.  
\par
The Mott-Littletone method is especially valuable to estimate 
key polaron energies because the long-range polarization fields 
are treated properly; many other methods 
(such as cluster methods or periodic cell methods) 
treat these significant terms badly.  

\section{Electronic holes in $\rm LaMnO_3$}  
\subsection{Relaxation energies of the localized holes in $\rm LaMnO_3$}

We study possible hole localisation (self-trapping) on 
Mn$^{3+}$ and on O$^{2-}$ ions in slightly hole-doped or 
non-stoichiometric LaMnO$_3$ crystal. Theoretical predictions of 
electron charge carrier self-trapping in ideal lattice are based on calculations 
of the so-called self-trapping energy \cite{Shluger & Stoneham
1993}, which is the 
difference between the localisation and relaxation energies. The 
first of these terms is basically an increase in the hole (electron) 
kinetic energy due to its localisation on a finite number of lattice 
sites from a completely delocalized state. The second is the energy 
gain due to the lattice polarisation by the localized charge. 
They represent a very delicate balance of large terms which in many 
cases differ by only 0.1 eV. The calculation of the localisation 
energy, especially in complex crystals, is the most difficult part 
of study of electron charge carrier self-trapping \cite{Shluger & Stoneham 1993}, 
and needs accurate electronic structure calculations not within  
the scope of this work. Our aim is rather to compare the relaxation 
energies for the hole localisation in two different sublattices of 
the same crystal. These energies are indicative of the strength of 
the electron-phonon interaction and their difference can suggest 
whether there are major differences in hole trapping in one of the 
sublattices.
\par
The process of the hole formation can be generally seen as the 
ionisation of the {\it in-crystal} ion with an electron being 
taken out of the crystal and put on the vacuum level. The energy 
required in this process (hole formation energy, $E^{\alpha}_h$, 
$\alpha$ = Mn, O, La) is the work done against the {\it in-crystal} 
ionic core potential, $I^{\alpha}$, and the crystalline electrostatic 
potential, $U^{\alpha}_{M}$, less then energy gain due to the lattice 
polarisation effects, $R^{\alpha}$    
\begin{eqnarray}
   E^{\alpha}_h = I^{\alpha} + U^{\alpha}_{M} + R^{\alpha}.
   \label{eqn2}
\end{eqnarray} 
To assess the extent of the lattice perturbation by the hole 
localisation and calculate the hole relaxation energy, it is useful 
to distinguish the 'electronic' and 'ionic' terms in the 
polarisation energy. The former term, which we will call 
$R^{\alpha}_{opt}$, is due to the 'electronic' polarisation of ions 
by the momentarily localized hole, which in our method is represented 
by the displacements of shells with respect to the cores which are 
fixed at their perfect crystal positions. It takes into account the
lattice response after e.g. Franck-Condon photoionisation. The 
lattice distortion term due to displacements of cores and related 
adjustment of shells after complete lattice relaxation, denoted as 
$\Delta R^{\alpha}_{th}$, is the difference between
the full polarisation energy, $R^{\alpha}$, and the $R^{\alpha}_{opt}$ 
\begin{eqnarray}
      \Delta R^{\alpha}_{th} = R^{\alpha} - R^{\alpha}_{opt}. 
      \label{eqn3}
\end{eqnarray}
It represents the hole relaxation energy. If this energy exceeds 
the localisation energy, i.e. the kinetic energy rise due to 
complete hole localisation on this site, then one can talk about 
the hole being self-trapped on this site. 
Given this assumption, Eq. (\ref{eqn2}) takes the form
\begin{eqnarray}
   E^{\alpha}_h = I^{\alpha} + U^{\alpha}_{M} + 
   R^{\alpha}_{opt} + \Delta R^{\alpha}_{th}.
   \label{eqn4}
\end{eqnarray} 
The shell-model Mott-Littleton calculations give the cumulative energy 
of the second and third term, $S^{\alpha}_{opt}$, or 
of the last three terms, $S^{\alpha}_{th}$, in Eq. (\ref{eqn4}) depending 
whether both shells and cores or shells only were allowed to relax. 
It is sensible, however, to evaluate these terms separately. This
can be rigorously done by calculating independently the on-site 
electrostatic potential $U^{\alpha}_M$ within the periodic model 
and using the definition introduced in Eq. (\ref{eqn3}). 
The values of $S^{\alpha}_{opt}$, $S^{\alpha}_{th}$ and the 
calculated terms of $U^{\alpha}_M$, $R^{\alpha}_{opt}$, and 
$\Delta R^{\alpha}_{th}$ are summarized in Table III.
\par
It follows from these calculations that there is a large difference 
in the lattice relaxation energies for O$^-$ and Mn$^{4+}$ 
holes. The lattice relaxation energy, -$\Delta R^{\alpha}_{th}$, 
caused by the hole localisation at the O-site (2.38 eV) appears to be 
significantly larger than that for the hole localized at the Mn-site 
(0.83 eV), as shown in Table III, (a). This indicates on the strong 
electron-phonon interaction in the case of the hole localized at the 
O-site and could suggest that the hole trapping is more preferential 
in the oxygen sublattice. 
However, the width of the Mn(3d) subband in the density of states, 
which determines the hole localisation energy, is much narrower 
than that of the O(2p) related subband \cite{Picket}. Therefore 
without a much fuller electronic structure calculation of the 
localisation energy it is impossible to draw any final conclusion 
as to in which sublattice the holes could be localized.
\par
One experimental test could involve the analysis of local 
vibrations due to the hole localisation. It can be facilitated by 
the qualitative difference in the lattice relaxation around the two 
centres which is clearly seen in Figures 2 and 3. The fully relaxed 
configuration of the ions surrounding the Mn$^{4+}$ electronic hole 
defect (see Figure 2) corresponds to the positions of cores in 
region I which have appreciable displacements ($\geq$ 0.004 \AA) 
from their perfect lattice sites. The cores of the six nearest 
neighbour oxygen ions are displaced symmetrically by about  0.1 \AA \ 
towards the Mn$^{4+}$ ion carrying the hole. The rest of the lattice 
relaxation comprises small displacements of Mn and La ions (of about 
0.01 and 0.004 \AA, respectively) out from the Mn$^{4+}$ hole 
center. 
\par
By contrast, the ionic relaxation around the O$^-$ hole center has 
the  axial symmetry,  with the largest lattice displacements of the 
nearby Mn ions (of about 0.21 \AA) along the axis away from the O$^-$ 
hole center (see Figure 3). These displacements cause the next two 
apical oxygen ions along the axis to move away from the O$^-$ hole 
center by about 0.1 \AA. The equatorial oxygen ions in the 
octahedron relax towards the hole center by about 0.03 \AA. 
In-plane La ions also show appreciable displacements away from the 
O$^-$ hole center. The qualitative difference in the symmetry of 
the lattice relaxation around the two centres implies the difference 
in the local vibrational modes, which can be used for experimental 
probing of hole localisation in LaMnO$_3$.

\subsection{Photoemission spectra and "{\it in-crystal}" ionisation potentials
in $\rm LaMnO_3$. Formation energies of the localized holes in $\rm LaMnO_3$ crystal} 

In order to evaluate the hole formation energy, we need to estimate the 
values of the unknown {\it in-crystal} ionisation energies, $I^{\alpha}$. 
We suggest estimating the ionisation potentials from the experimental 
photoemission spectroscopy (PES) data, which can be directly related to our 
calculations. PES at different excitation energies 
probes in principle bonding states as well as non-bonding states. 
The latter, being ion-in-crystal-like, can be related to the 
Frank-Condon energies obtained in our calculations. In order to 
juxtapose experimental and calculated values we need also to take 
into account that PES binding energy, $E_{PES}$, is measured with respect to 
sample's Fermi energy level $E_F$. So, we write
\begin{eqnarray}
I^{\alpha} + U^{\alpha}_{M} + R^{\alpha}_{opt} = E^{\alpha}_{PES} + E_F.
\label{eqn5}
\end{eqnarray}
In the PES spectra of LaMnO$_3$ there are two main photoemission bands 
around 3.5 and 6 eV binding energies at T = 100, 200 K for the He I 
($h\nu$ = 21.2 eV) and He II ($h\nu$ = 40.8 eV) photon energies 
for which the O(2p) photoionisation cross-section is dominant 
\cite{Saitoh}. The main maximum at 3.5 eV has been assigned 
primarily to the O(2p) non-bonding states, whereas the second 
maximum is assigned to the Mn(3d) - O(2p) bonding states and the 
decreasing of O(2p) character correlates with decreasing of 
Mn(3d) - O(2p) hybridization strength. For higher energies with He II 
PES study the Mn(3d)/O(2p) cross-section ratio increases 
and a feature near 2.7 eV  appears \cite{Saitoh}. At high photon 
energies, 500 eV, and T = 280 K, the band at 3.5 eV is not clear 
evident, but the band at 2.7 eV becomes dominant over the band at 
6 eV, which stands for maximum contribution from Mn(3d) 3t$_{2g}$ 
states at 2.7 eV binding energy \cite{Park}. The crystal field (CF) 
splitting between Mn(3d) 3t$_{2g}$ and e$_g$ states in LaMnO$_3$ 
has been estimated to be about $\Delta_{CF}$ $\simeq$ 1.5 eV from the PES 
study \cite{Park}. The peak at 17 eV has been assigned to the La(5p) 
states \cite{Park}. 
\par
So, in accordance with the dominant contributions to the PES spectra of LaMnO$_3$ 
\cite{Saitoh,Park}, we assign the following values
$E^{O}_{PES} \simeq$ 3.5 eV, $E^{La}_{PES} \simeq$ 17.0 eV, and  
$E^{Mn}_{PES} \simeq$ 1.2 eV, suggesting the process of Mn hole formation is 
associated with electron photoionisation from e$_g$ level.       
These maxima in the PES spectra correlate well with the maxima in the density of 
states (DOS) for O(2p) and Mn(3d) e$_g$ valence bands in LaMnO$_3$ 
calculated within the local spin density approximation (LSDA) \cite{Picket}. 
The corresponding schematic representation of the band structure in 
accordance with the assigned maxima of binding energies in the 
PES spectra \cite{Saitoh,Park} in the scale of energies related to the 
crystal Fermi level $E_F$ is shown in Figure 4. The gap in the 
e$_g$ electron band opened at the E$_F$ due to the lattice 
distortion (the Jahn-Teller effect and/or lattice mismatch effect) 
is shown in accordance with the CF splitting PES data \cite{Park}.
The relevant electron excitations from the Mn(3d) e$_g$, O(2p) and 
La(5p) valence band levels are schematically shown by arrows. The 
corresponding PES energies, $E^{\alpha}_{PES}$, are summarized in Table III.
\par
Having assigned the $E^{\alpha}_{PES}$ energies we now proceed with the 
evaluation of the hole formation energies $E^{\alpha}_{h}$. First, we 
obtain the crystal Fermi energy by using Eq. (\ref{eqn5}) and data for 
the La ion. We assume that the electronic
density of the closed-shell La$^{3+}$ ion is not significantly deformed by the crystalline 
field, so {\it in-crystal} ionisation energy, $I^{La}$, can be 
plausibly estimated by the forth standard ionisation potential of a 
free La atom \cite{Handbook}, presented in Table III. This approximation is consistent 
with the full ionic charges adopted in our shell model parameterization. 
We note that the approximation of a free cation just made, is shown 
to be reliable only for the closed-shell cations. This gives 
$E_F \simeq$ 1.36 eV for the Fermi energy of LaMnO$_3$ crystal.
\par
The situation is more complicated in the case of manganese and oxygen.
The Mn$^{3+}$ ion has non-closed 3d shell with four electrons in it,    
so we expect that the {\it in-crystal} ionisation energy, $I^{Mn}$,
could differ from the fourth ionisation potential of a free Mn atom.
The O$^{2-}$ ion is only stabilized by the crystalline field, thus it 
has a negative ionisation potential which can not be defined in a 
non-speculative way. Using the Mn(3d) and O(2p) related maxima in the 
PES spectra, $E^{\alpha}_{PES}$, and the obtained value for $E_F \simeq$ 1.36 eV, 
we can now estimate the effective ionisation energies, $I^{\alpha}$, 
for manganese and oxygen in LaMnO$_3$ crystal from Eq. (\ref{eqn5}). 
These values are presented in Table III, the free metal ionisation 
potentials \cite{Handbook} are given for comparison in brackets. The O$^{2-}$ 
{\it in-crystal} ionisation potential $I^O$ (negative electron 
affinity of O$^-$) is estimated thereby to be -13.91 eV. This 
absolute value is within the limits of O$^-$ electron affinities 
calculated for many oxide compounds in ref. \cite{Harding} using an 
embedded cluster {\it ab-initio} method. Those calculations 
predicted 10.6 eV for MgO and 12.9 eV for ThO$_2$. Taking into 
account the semi-empirical nature of our calculations we find 
this agreement quite good. 
\par
The optical and thermal energies of hole formation,
$E^{\alpha}_h(opt)$ and $E^{\alpha}_h(th)$, are calculated using 
these effective values of the {\it in-crystal} ionisation energies 
in accordance with Eq. (\ref{eqn4}) and presented in Table III.
Taking into account the CF splitting effect we have found out that 
the electronic hole is marginally more stable at the Mn-site than 
at the O-site in the LaMnO$_3$ lattice, but the energy difference 
between the thermal energies of the hole formation, 
$E^{\alpha}_h(th)$, is too small (0.75 eV). This result 
rather suggests that both possibilities should be treated seriously.
That is, providing the balance between the localisation and relaxation
energies favours the possibilities for the hole self-trapping at 
the Mn- and O-sites, the electronic hole in LaMnO$_3$ will be 
likely localized on the manganese, or on both oxygen anion and 
transition metal cation, rather then on the oxygen ion alone.   
\par
In order to assess the accuracy of the calculated energies of the hole 
formation and lattice relaxation we need to discuss the following issue. 
This issue concerns the pair potentials used in these calculations. 
The energies presented in Table III, (a) were obtained using the pair 
potentials listed in Table I, (a). To check the robustness of our 
results, we repeated the same calculations using the potentials 
from R. Grimes and D. Bradfield \cite{Robin}, which give close 
values for the dielectric constants in LaMnO$_3$ (see Table II, (c)), 
but were specially optimized to treat different Mn$^{3+}$ and Mn$^{4+}$ 
charge states. The calculated values of formation and 
polarisation energies for the localized holes Mn$^{4+}$, O$^-$ and 
La$^{4+}$ and the energies deduced in Eq. (\ref{eqn4})
using these pair-potentials are presented in Table III, (c). 
These calculations demonstrated that the hole 
relaxation energy of Mn$^{4+}$ is decreased by 0.16 eV if we account 
for the change in the short-range potentials caused by the change 
of the Mn charge state. Comparing with our results, 
we can see good coincidence for the similar values and 
for the thermal and optical energies of hole formation.

\section{Optical charge-transfer transitions in $\rm LaMnO_3$}

Polaronic-type electron charge carriers mostly determine specific transport 
properties of CMR materials in their high-temperature insulating paramagnetic phase, 
which are always associated with photo-induced charge-transfer (CT) transitions. 
In hole-doped systems of perovskite manganites of R$_{1-x}$A$_x$MnO$_3$ 
the main important CT transitions associated with localized charge carriers
will be apparently those involving Mn$^{4+}$ and O$^-$ self-trapped holes.
In this section using the derived values of the {\it in-crystal} 
ionisation energies we calculate energies of the main CT transitions 
suggesting that holes could be localized at the Mn- or O-sites. 
We analyze the contribution of these 
CT transitions to the experimental optical conductivity in non-stoichiometric 
or slightly hole-doped LaMnO$_3$ crystals to make the assignment of the bands 
in the optical conductivity spectrum more clear-cut and to verify our 
shell model approach. Now we proceed with brief analysis of the optical
conductivity spectra.  

\subsection{Analysis of the optical conductivity spectra in $\rm LaMnO_3$}

The room temperature optical conductivity spectrum of LaMnO$_3$ measured  
in \cite{Arima and Tokura} is shown by solid curve 1 in Figure 5, (a) in the 
spectral region 0 to 8 eV (reproduced from the original data with a 
permission of T. Arima and Y. Tokura \cite{Arima and Tokura,Private}). 
This spectrum is very similar to that measured by Okimoto {\it et al.} 
at T = 9 K \cite{Okimoto}. It reveals the optical gap near 1.3 eV and 
includes several broad absorption bands with maxima at $\sim$ 2.3, 5 and 9 eV. 
The gap is assumed to be of the charge-transfer type \cite{Arima and Tokura}. 
The first transition at $\sim$ 2.3 eV has been suggested to be of 
the O(2p) - Mn(3d) character. The band at $\sim$ 5 eV is thought to be due 
to the excitations to a higher-lying Mn 3d e$_g$ antiparallel spin configuration, 
separated by a Hund's rule coupling energy. 
The wide band observed around 9 eV in the optical conductivity spectrum is assigned 
to the O(2p) - La(5d) interband optical transition \cite{Arima and Tokura}. 
\par
The optical spectra measured in hole-doped manganese oxides 
show striking changes over a wide photon region (0 to 6 eV) as the temperature 
and doping concentration change. In La$_{1-x}$Sr$_x$MnO$_3$ system, with  
increasing doping concentration (x = 0 to 0.3, T = 9 K, \cite{Okimoto}), the 
excitations around 2.3 and 5 eV shift appreciably to lower energies. 
However, the principal changes take place in the low-energy mid-infrared  
spectral region stemming from the filling of the gap because of the hole 
doping. In the insulating paramagnetic phase of hole-doped manganites
there are two features clearly observed in the experimental mid-infrared 
optical conductivity, 
around 0.6 eV \cite{Okimoto,Boris1} and around 1.2 - 1.5 eV \cite{Lee,Mayr}. 
The optical band at $\sim$ 0.6 eV seems 
to be associated with polaronic-type charge carriers 
in doped CMR manganites, and the consistent value of activation energy 
of $\sim$ 0.15 eV was measured for the hopping conductivity 
in the adiabatic temperature limit \cite{Jaime,Machida}. 
The origin of these features is still a subject of 
many controversial discussions. It is well known that LaMnO$_3$ crystal
has a strongly distorted orthorhombic structure at low temperatures, which
in many works is ascribed due to strong electron-phonon interaction stemming 
from Jahn-Teller (JT) effect inherent for Mn$^{3+}$ ion in the octahedral 
oxygen configuration. In this case the $e_g$ bands split into two subbands, 
separated by the Jahn-Teller energy, $E_{JT}$. As the on-site $d - d$ transitions
are dipole-forbidden, these mid-infrared peaks around 0.6 and 1.2 - 1.5 eV were
qualitatively explained as due to an electron transition from an occupied
site Mn$^{3+}$ to an unoccupied site Mn$^{4+}$ and to an adjacent occupied 
site Mn$^{3+}$, respectively \cite{Millis}.
\par
In a recent theoretical study of the optical conductivity spectra of 3d 
transition metal perovskites LaMO$_3$ (M = Ti - Cu) \cite{Solovyev} using the 
local spin density approximation (LSDA + U) method, the authors estimated 
the role of the lattice distortions in the band structure calculations and 
concluded that the Jahn-Teller structural distortions play a crucial role in 
opening the optical gap in LaMnO$_3$ Mn(3d) e$_g$ valence band. Considering the 
experimentally observed distorted structure of LaMnO$_3$ crystal, the direct 
gap in the LSDA study has been estimated to be $\approx$ 0.7 eV, which is 
less than the observed optical gap ($\approx$ 1.3 eV \cite{Arima and Tokura,Okimoto}). 
There are also some discrepancies observed at higher energies between the 
experimental optical conductivity in the 3d transition metal perovskites of 
LaMO$_3$ and the calculated optical conductivity considering contributions from 
the interband and intraband transitions for the perfect lattice \cite{Solovyev}, 
which makes the assignment of the optical bands complicated. 
In addition, the contribution from the CT transitions to the optical conductivity 
in the non-stoichiometric lattice must be taken into account to describe 
satisfactorily the optical conductivity at low energies and to make the assignment 
of the optical bands in LaMnO$_3$ crystal more clear-cut.   
\par
To estimate the contribution to the experimental optical conductivity of
LaMnO$_3$ crystal, shown by solid curve 1 in Figure 5, (a) from the CT transitions 
we have analyzed the imaginary part of the dielectric function, $\epsilon_2(\nu)$ 
\cite{Arima and Tokura,Private}. For this purpose, we presented the $\epsilon_2(\nu)$ 
spectrum, shown by solid curve 1 in Figure 5, (b), by a sum of the first 
three main bands with Lorentzian lineshapes 
\begin{eqnarray} 
\epsilon_2(\nu) = \sum \nu_{{\bf \rm p}i}^2 \gamma_i \nu / [(E_i^2 - \nu^2)^2 + \gamma_i^2\nu^2]
\label{eqn6}
\end{eqnarray}     
where ($\nu_{{\bf \rm p}i}/E_i)^2$ = $f_i$ is the oscillator strength, 
$\gamma_i$ is the width of the band and $E_i$ is the resonance frequency of 
the {\it i}-oscillator. The three Lorentzian bands with the maxima $E_i$ at 
1.93, 4.75 and 9.07 eV and widths $\gamma_i$ of 1.46, 2.0 and 5.1 eV, 
respectively, are represented by short-dashed 2, dotted 3 and long-dashed 4
curves in Figure 5, (b). The rest of the imaginary part of the dielectric function 
after subtraction of the Lorentzian bands is shown by light curve 5. 
The Lorentzian band parameters together with the estimated oscillator strengths
are given in Table IV. These Lorentzian bands contribute to the experimental 
optical conductivity spectrum, as shown by the corresponding lines in 
Figure 5, (a), with more details at low energies.  

\subsection{Calculation of charge-transfer transition energies}

Using a Born-Haber cycle and the shell model we can consider both thermally 
assisted and optical CT processes. This can be illustrated for a 
hypothetical case of two ions X$^{(m+1)+}$, Y$^{(n-1)+}$  
transformation into X$^{m+}$, Y$^{n+}$ in which an electron is transferred 
from Y to X (or a hole from X to Y):
\begin{equation}
X^{(m+1)+} + Y^{(n-1)+} \Rightarrow X^{m+} + Y^{n+}. 
\end{equation}
There are two basic steps: 
1. Remove an electron from the {\it in-crystal} Y$^{(n-1)+}$ ion  
to infinity, outside the crystal. 2. Add an electron from the infinity, outside 
the crystal, to the {\it in-crystal} X$^{(m+1)+}$ ion.
The steps are standard within the shell model. Whether or not shells alone, or 
shells and cores, are relaxed depends on which transition is being calculated. 
In the case of thermally assisted hopping, the shell and core positions are 
considered to be fully relaxed in both charge states and the transition energy is 
denoted $E_{th}$. Comparison of the two charge states gives an additional indication 
which species are more stable. For optical transitions, the Franck-Condon 
approximation is used and their energies $E_{opt}$ are calculated on the 
assumption that only shells are able to relax (corresponding to full electronic 
polarisation), whereas the cores remain in the positions corresponding to the 
initial state. 
The major contributions into these energies come from ionisation energies, 
$I^Y_n$, $I^X_{(m+1)}$, and the Madelung and polarisation 
terms, which cumulative energies for defect configuration corresponding the CT
transition considered, $S[X^{m+},Y^{n+}]_{(opt,th)}$, result 
from the Mott-Littleton calculations (like in Eq. \ref{eqn4}). 
If the CT includes the localized hole in 
thermal equilibrium in the initial state (the related values in LaMnO$_3$ 
lattice are presented as $S^{\alpha}_{th}$ in Table III), 
the correspondent  thermal energy for initial defect configuration, 
$S[X^{(m+1)+},Y^{(n-1)+}]_{th}$, must be subtracted. Thus, the thermal
and optical energies of the CT transitions can be calculated using the
following formula
\begin{eqnarray}
    E_{opt} = I^Y_n - I^X_{m+1} + S[X^{m+},Y^{n+}]_{opt} -
    S[X^{(m+1)+},Y^{(n-1)+}]_{th} \\
    E_{th} = I^Y_n - I^X_{m+1} + S[X^{m+},Y^{n+}]_{th} -
    S[X^{(m+1)+},Y^{(n-1)+}]_{th}.
\end{eqnarray}
There will clearly be some dependence on the separation of X and Y. The CT 
optical transitions for nearest neighbours are likely to dominate and the 
relevant key cases have been calculated. If X and Y are the same (symmetric) 
the ionisation terms cancel out, as for intervalence CT transition 
Mn$^{4+}$ + Mn$^{3+}$ $\Rightarrow$ Mn$^{3+}$ + Mn$^{4+}$. We would like to 
emphasize here that the calculations of CT transitions between the ions of 
metal Mn sublattice are more reliable because they do not depend on the 
difference in the Madelung potential between the two sublattices, nor on the 
phenomenologically deduced parameter of the O$^{2-}$ {\it in-crystal}  
ionisation potential. 
\par
The cumulative thermal and optical energies following from the
Mott-Littleton calculations, $S_{th}$ and $S_{opt}$, 
for the CT transitions involving Mn$^{4+}$ and O$^{-}$species, 
and those characterizing fundamental electronic transitions in LaMnO$_3$ lattice, 
such as, Mn(3d) gap transition, O(2p) - Mn(3d), and O(2p) - La(5d) 
are presented in Table V by the transitions 1-3 and 4-6, respectively. 
To calculate the optical and thermal energies of the CT transitions 
we used a self-consistent set of the ionisation potentials
(see Table III, (a)), derived by matching the calculated optical energies of the hole 
formation with the photoemission experimental energies and the standard ionisation 
potentials for a La free atom. We also need to estimate 
the third {\it in-crystal} ionisation potential of Mn, 
$I^{Mn}_{III}$. We suggest that it should be shifted {\it in-crystal} from 
the standard value for a Mn free atom (33.67 eV \cite{Handbook}) 
by the same value as the 
fourth potential of Mn (from the standard value 51.2 eV), by subtraction the 
crystal field (CF) splitting effect ($\Delta_{CF} \simeq$ 1.5 eV), so we calculate: 
{\it in-crystal} $I^{Mn}_{III}$ = 33.67 - (51.2 - (47.41 + $\Delta_{CF}$/2)) 
= 30.63 eV. 
\par
Taking the standard value of $I^{La}_{III}$ = 19.18 eV, 
the calculated optical energy of the fundamental transition 
of CT character O(2p) - La(5d), 
$E_{opt} = I^O - I^{La}_{III} + S[La^{3+},O^{2-}]_{opt}$ = 8.93 eV 
(see Table V, transition 6) 
correlates well with the maximum of the broad band in the $\epsilon_2$ function 
near 9.07 eV (long-dashed curve 4 in Figure 5, (b)). This encouraging consistency 
between the experimental and calculated energies let us to suggest that 
the earlier estimated value of $I^O$ = -13.91 eV {\it in-crystal} provides 
a reasonable value in this shell model calculations. We also calculated optical 
energy of the fundamental transition of CT character O(2p) - Mn(3d), 
$E_{opt} = I^O - I^{Mn}_{III} + S[Mn^{3+},O^{2-}]_{opt}$ 
= 5.61 eV (see Table V, transition 5). We suggest that the relevant transition 
should correlate with the broad optical band observed in the $\epsilon_2$ 
function near 4.75 eV (dotted curve 3 in Figure 5, (b)).  
Our calculations therefore predict transitions which appear to correlate
with the maxima of the major broad-band features in the optical conductivity spectrum. 
The calculated optical energy for the transition between the Mn(3d) 
valence band and the upper Hubbard Mn(3d) band is estimated to be 
$E_{opt} = I^{Mn}_{IV} - I^{Mn}_{III} + S[Mn^{3+}, Mn^{3+}]_{opt}$ = 
3.72 eV (see  Table V, transition 4) predicting 
the Mott-Hubbard band gap type in LaMnO$_3$ crystal.
This value agrees well with the assigned transition experimentally observed 
at $\sim$ 3.5 eV in Nd$_{0.7}$Sr$_{0.3}$MnO$_3$ \cite{Lee} and at $\sim$ 
3.2 eV in La$_{0.825}$Sr$_{0.175}$MnO$_3$ \cite{Takenaka}. 
A small contribution to the experimental optical conductivity
can be observed around 3.7 eV in LaMnO$_3$ crystal, 
as shown by light curve 5 in Figure 5, (b) resulting from our dispersion analysis.
\par
Having assigned the fundamental electronic transitions in 
LaMnO$_3$ crystal in accordance with the results of our calculations, which are 
also consistent with the consideration given in ref. \cite{Lee}, we should note 
that the assignment of the optical conductivity band at $\sim$ 2.3 eV still remains 
controversial. In the earlier study \cite{Arima and Tokura} 
this band has been associated with the fundamental CT transition of 
O(2p) - Mn(3d) e$_g$ character, whereas the band at about 5 eV has been associated by the 
authors with the excitations to a higher-lying Mn 3d e$_g$ antiparallel spin 
configuration, separated by a Hund's rule coupling energy.  
However, our results allow us to argue that an alternative interpretation of 
this transition obtained in this work can be correct. We suggest that the band 
at $\sim$ 2.3 eV is rather associated with the presence of 
Mn$^{4+}$ and/or O$^-$ localized holes in LaMnO$_3$ crystal, 
which is known to exhibit strongly non-stoichiometric 
behavior with respect to oxygen content, up to 0.1 in as-grown crystal.
\par 
Indeed, if an optical band is associated with a CT transition 
in a crystal lattice, its maximum position, $h \nu_{max}$, and the half-width, 
$\Delta W$, are known to be linked by a simple formula 
in the high temperature limit \cite{Hush} 
\begin{eqnarray} 
h \nu_{max} = (16 \cdot (ln2) kT)^{-1} {\Delta W^2}.   
\label{eqn10}
\end{eqnarray} 
We can invoke this expression to verify the CT transition character of the bands
associated with photo-induced hopping conductivity of the localized charge carriers.
Using this expression, estimates for T = 300 K show very 
encouraging consistency between the half-width and the maximum energy of 
the first Lorentzian band (short-dashed curve 2 in Figure 5, (b), with the parameters 
given in Table IV): from $\Delta W \simeq$ 0.73 eV we obtain 
$h \nu_{max} \simeq$ 1.92 eV, which matches well the maximum position
estimated to be near 1.93 eV from the dispersion analysis of the $\epsilon_2$ 
function. So, this is consistent with the view that this transitions could be 
of CT-type, associated with the presence of localized electronic charge carriers 
in LaMnO$_3$ crystal lattice. 
\par
Among these main contributions to the optical 
conductivity will be expected from the following CT transitions: 
1) the intervalence Mn$^{3+}$/Mn$^{4+}$ CT transition, 
Mn$^{4+}$ + Mn$^{3+}$ $\Rightarrow$ Mn$^{3+}$ + Mn$^{4+}$, 
$E_{opt} = S[Mn^{3+},Mn^{4+}]_{opt} - S^{Mn}_{th}$ = 1.33 eV,     
2) the transition of the O$^-$ self-trapped hole to a neighbouring manganese ion, 
O$^{-}$ + Mn$^{3+}$ $\Rightarrow$ O$^{2-}$ + Mn$^{4+}$,
$E_{opt} = I^{Mn}_{IV} - I^O + S[O^-,Mn^{3+}]_{opt} - S^O_{th}$ = 1.43 eV,  
3) the transition of the Mn$^{4+}$ self-trapped hole to a neighbouring oxygen ion, 
Mn$^{4+}$ + O$^{2-}$ $\Rightarrow$ Mn$^{3+}$ + O$^{-}$, 
$E_{opt} = I^O - I^{Mn}_{IV} + S[O^{2-},Mn^{4+}]_{opt} - S^{Mn}_{th}$ = 2.98 eV, 
(see transitions 1 - 3, respectively, in Table V).
\par 
Analyzing all calculated and experimental optical energies given in Table V
we can conclude that the agreement is much better for those calculations 
not including the {\it in-crystal} ionisation potentials of manganese, 
$I^{Mn}_{III}$ and $I^{Mn}_{IV}$, or in the case when their difference enters
and the inaccuracy due to these terms cancels out. Relying on the correlation
between the calculated and experimental optical energies we can try to 
refine the values of $I^{Mn}_{III}$ and $I^{Mn}_{IV}$, which determination
{\it in-crystal} presents difficulties due to non-closed 3d shell of Mn$^{3+}$
ion. Indeed, expecting correlation between the calculated optical energy
of the fundamental transition of CT character O(2p) - Mn(3d), 
$E_{opt} = I^O - I^{Mn}_{III} + S[Mn^{3+},O^{2-}]_{opt}$ = -13.91 - 30.63 + 50.15 
= 5.61 eV (see Table V, transition 5) with the broad optical band observed 
in the $\epsilon_2$ function near 4.75 eV (dotted curve 3 in Figure 5, (b)) 
we can refine the third {\it in-crystal} ionization potential of manganese 
as $I^{Mn*}_{III}$ = 31.49 eV and, correspondingly, $I^{Mn*}_{IV}$ = 48.27 eV.
Using these corrected values we recalculated the energies of 
transitions 2 and 3 in Table V associated with the CT transitions of O$^-$ 
and Mn$^{4+}$ self-trapped holes and obtained close values of optical
energies $E_{opt}$ = 2.29 eV and $E_{opt}$ = 2.12 eV, respectively. 
These corrected values for the optical CT transition energies are presented 
in brackets in Table V for transitions 2 and 3.
We suggest that these CT transitions 
Mn$^{4+}$ + O$^{2-}$ $\Rightarrow$ Mn$^{3+}$ + O$^{-}$ and 
O$^{-}$ + Mn$^{3+}$ $\Rightarrow$ O$^{2-}$ + Mn$^{4+}$  
associated with the hole transfer along a chain 
Mn$^{4+}$ - O$^{2-}$ - Mn$^{3+}$ could be responsible for 
the band around 2.3 eV in the optical conductivity spectrum 
(the related band in the $\epsilon_2$ spectrum has maximum energy 1.93 eV)
of as-grown non-stoichiometric LaMnO$_3$ crystal. 
If the band is assigned so, the net oscillator strength of this band 
$f_i$ = 0.51 (see Table IV) will be dependent on the concentration of 
the localized holes as $f_i = f_{CT}/x$ providing an estimate 
of oscillator strength for the CT transition, $f_{CT}$. The typical value 
of $x \sim$ 0.1 for as-grown LaMnO$_3$ crystal gives the estimate 
for the oscillator strength consistent with the transition of CT-type.       
\par
The negative value of the thermal energy, $E_{th}$ = -0.75 eV, 
for transition 2 from Table V indicates on more thermally stable state
of the Mn$^{4+}$ hole compared to the O$^-$ hole state, in accordance with 
our results for the thermal energies of the holes formation $E^{\alpha}_{th}$ 
(Table III, (a)) based on preliminary estimates for the fourth {\it in-crystal} 
ionization potential of manganese ion. 
Using now the refined value of $I^{Mn}_{IV}$ = 48.27 eV, deduced from 
the comparison between the calculated and experimental optical energies, 
we derive the thermal energies for transitions 2 and 3 of 0.12 and -0.12 eV, 
respectively. This result reinforces our arguments made above 
that the electronic hole can be thermally stable on both 
transition metal cation and oxygen anion in LaMnO$_3$ crystal. 
\par
In accordance with our shell model calculations the intervalence CT transition 
Mn$^{4+}$ + Mn$^{3+}$ $\Rightarrow$ Mn$^{3+}$ + Mn$^{4+}$ is predicted to 
have the optical energy $E_{opt}$ = 1.33 eV (Table V, transition 1), compared
with the energy of the optical gap in LaMnO$_3$, and it is not observable 
in as-grown pure crystal.
\par
Due to doping effect the optical spectra in CMR manganese oxides 
show striking changes over a wide photon region (0 to 6 eV). 
In La$_{1-x}$Sr$_x$MnO$_3$ system, with  
increasing doping concentration from x = 0 to 0.3 at T = 9 K \cite{Okimoto}, 
the optical conductivity bands around 2.3 and 5 eV shift by more than 0.5 eV 
to lower energies. We have analyzed the low-energy $\epsilon_2$ function 
in a slightly doped La$_{7/8}$Sr$_{1/8}$MnO$_3$ compound \cite{Mayr},
and found it to be well described by the Lorentzian with the maximum 
at 1.32 eV and the half-width of 0.61 eV, as presented in Figure 6. 
We would like to emphasize that the maximum position of this band and
its half-width are also in a good correlation with the formula describing
transition of CT character (see Eq. \ref{eqn10}): 
from $\Delta W \simeq$ 0.61 eV we obtain $h \nu_{max} \simeq$ 1.34 eV, 
which matches well the maximum position experimentally observed.  
It would appear reasonable to suggest that this band is due to the same
origin as the band at $\epsilon_2$ = 1.93 eV in pure LaMnO$_3$ compound, 
assigned due to transitions 2 and 3, Table V, and shifted by about 0.5 eV 
to lower energies due to the hole interaction effect in CMR systems. 
Using this line of reasoning,
we can also suggest that the 0.6 eV band \cite {Okimoto,Boris1} in optical
conductivity of CMR compounds is due to the intervalence CT transition 
Mn$^{4+}$ + Mn$^{3+}$ $\Rightarrow$ Mn$^{3+}$ + Mn$^{4+}$, 
and associated with photo-induced hopping conductivity 
of Mn$^{4+}$ localized holes, with the consistent value of 
hopping conductivity activation energy of $\sim$ 0.15 eV measured 
in the adiabatic temperature limit \cite{Jaime,Machida}. 
\par 
The results given are based on the shell model parameters (Table I,(a)), 
which were fitted to give good values for both the dielectric 
constants and the transverse optic modes. When we use the second set of 
the shell model parameters, determined primarily using the oxide structures MnO, 
LaMnO$_3$, and Ca$_2$MnO$_4$ \cite{Robin} (Table 1, (c)), 
the resulting energies are very similar for the low energy optical 
CT transition band near 2.3 eV, but the predicted energies are about 1.5 eV 
higher for $\sim$ 5 eV optical band. 

\section{Conclusions}

In this paper we explore the role of electronic and ionic polarisation energies 
in the physics of the CMR materials. In particular, 
we examine energies associated with localized holes of  Mn$^{4+}$ and O$^-$ in the 
lattice of the "parent" LaMnO$_3$ compound. Our calculations are made for 
the idealized cubic perovskite LaMnO$_3$ structure, which is relevant to 
the non-magnetic quasicubic perovskite structure experimentally observed at 
high temperatures T $\geq$ 400 K $>$ T$_N$ $\simeq$ 140 K. 
To estimate the polarisation energy terms we use a fully-ionic shell model. 
The shell model parameters we derive satisfy the equilibrium conditions for 
the quasicubic perovskite structure of LaMnO$_3$ and agree well with 
experimental values of the static and high-frequency dielectric constants 
as well as transverse optical phonons. 
\par
As a result of our shell model calculations we find that, from one side, 
there is a huge difference between the hole relaxation energies on the oxygen and 
manganese sites which indicates on the strong electron-phonon interaction in the case 
of the hole localized at the O-site.
From the other side, the difference, which we find 
between the thermal energies of Mn$^{4+}$ and O$^-$ holes is too small. 
This means, in fact, we should consider seriously the likelihood that 
the electronic hole in LaMnO$_3$ is localized on the manganese, or on both 
oxygen anion and transition metal cation, rather than on the oxygen ion
alone. If so, this system would be like many other transition metal oxides.
\par 
Assuming that holes in LaMnO$_3$ crystal can localize in either 
or in both Mn and O sublattices, we estimate the main associated optical 
CT transition energies, which we relate to the experimentally observed 
optical conductivity spectra. Applying the Mott-Littleton approach 
we estimate the CT transition energies within a Born-Hyber cycle 
using the {\it in-crystal} ionization potentials for the ions in LaMnO$_3$
crystal obtained in our consideration of the experimental photoemission spectra.  
\par
Our analysis allows us to suggest a new interpretation of the main bands in 
the optical conductivity spectrum at $\sim$ 2.3, 5 eV. We suggest that the band 
at $\sim$ 5 eV is associated with the fundamental O(2p)-Mn(3d) transition 
of CT character, whereas the band at $\sim$ 2.3 eV is rather associated with 
the presence of Mn$^{4+}$ and/or O$^-$ self-trapped holes in non-stoichiometric 
LaMnO$_3$ compound. 
\par 
To summarize, we believe that the results of this work demonstrate the 
applicability and usefulness of the shell model approach to preliminary 
modeling of polaron-related features in complex oxides such as CMR materials, 
and hope that they will stimulate further theoretical and experimental studies 
of the character and properties of hole states in these materials.

\begin{acknowledgments} 
The authors thank Dr. J. Gale for making available General Utility Lattice 
Program (GULP) used in the present calculations. 
We are greatly appreciate Y. Tokura and T. Arima for providing us the original
reflectivity spectra in LaMnO$_3$. We thank to F. Mayr and coauthors 
for the permission to reproduce their experimental data. 
We woul d also like to thank W.C. Mackrodt  
and A. Ionov for useful information. We are grateful to R.W. Grimes and 
D.J. Bradfield for fruitful discussions and for making available to us 
one set of interatomic potentials. 
We are grateful for Royal Society/NATO support of the visit 
to University College London of one of us (N. Kovaleva). 
\end{acknowledgments}

\begin{table} 
\caption
{Potential parameters for short-range interactions in LaMnO$_3$ 
({\it Pm3m}). (a) - elaborated in the present work;
(b) - from M.S. Islam {\it et al.} \cite{Islam};
(c) - from R. Grimes \cite{Robin} for Mn$^{3+}$ and Mn$^{4+}$ different 
valent states; $r_{cutoff}$ = 20\AA.}
\begin{ruledtabular}
\begin{tabular}{rccccc}    
   & A(eV) & $\rho $(\AA) & C(eV$\cdot $\AA $^{-6}$) & Y($\mid e \mid$) & k (eV 
$\cdot $\AA  $^{-2}$) \\    
\colrule
(a) La$^{3+}$:O$^{2-}$ & 1516.3 & 0.3639 & 0.00 & & \\
      Mn$^{3+}$:O$^{2-}$ & 1235.9 & 0.31525 & 0.00 & & \\ 
      O$^{2-}$:O$^{2-}$ & 22764.3 & 0.1490  & 20.37 & -2.48 & 16.8 \\ 
\colrule      
(b) La$^{3+}$:O$^{2-}$ & 1516.3 & 0.3525 & 0.00 & & \\ 
      Mn$^{3+}$:O$^{2-}$ & 1235.9 & 0.3281 &0.00 & &\\ 
      O$^{2-}$:O$^{2-}$ & 22764.3 & 0.1490 & 43.00 & & \\   
\colrule
(c) La$^{3+}$:O$^{2-}$ &  2088.79 & 0.3460  & 23.25 & & \\
      Mn$^{3+}$:O$^{2-}$ & 922.83 & 0.3389 & 0.00 &  & \\
      Mn$^{4+}$:O$^{2-}$ & 1386.14 & 0.3140 & 0.00 & & \\
      O$^{2-}$:O$^{2-}$  & 9547.96 & .2192 & 32.00 & -2.04 & 6.3 \\ 
\end{tabular}
\end{ruledtabular} 
\end{table} 

\begin{table} 
\caption{Crystal properties of LaMnO$_3$ ({\it Pm3m}) calculated using the shell 
model potentials (Table I) and compared with experimental~ data.}  
\begin{ruledtabular}
\begin{tabular}{lcclllll}  
$\ $ & Lattice  & Cohesive& $\epsilon_0$ &  
$\epsilon_{\infty}$ & $\omega_{TO_1}$   & $\omega_{TO_2}$     
& $\omega_{TO_3}$   \\  \cline{6-8} 
$\ $ &  const. a$_0$, (\AA)& energy E$_{lat}, $(eV) &  &  &  & ($cm^{-1}$) & 
 \\   
\colrule
Exp. & 3.889 &  $\ $ & 18 $\pm $ 2 \cite{Arima and Tokura} & 4.9 \cite{Arima and 
Tokura} & 
172 \cite{Arima and Tokura}& 360 \cite{Arima and Tokura}&  
560 \cite{Arima and Tokura}\\ 
Calc. (a) & 3.889 &  -140.52 & 15.6 & 4.9 & 172 & 308 & 513 \\   
Calc. (b) & 3.904 &  -139.12 & 56.17 &  &  &  &  \\ 
Calc. (c) & 3.906 &  -139.58 & 14.1  &   4.6 & 156 & 252 & 368 \\
\end{tabular} 
\end{ruledtabular}
\end{table}

\newpage

\begin{table}
\caption
{Formation and polarisation energies for localized holes in LaMnO$_3$:
(a) - for the pair potentials explored in this work; (c) - for the pair potentials 
from R. Grimes \cite{Robin} for Mn$^{3+}$ and Mn$^{4+}$ different 
valent states.} 
\begin{ruledtabular}
\begin{tabular}{crrlrrrrrr}   
$\alpha$-hole  & $E^{\alpha}_h(opt)$ & $E^{\alpha}_h(th)$ & \ \ \ \ $I^{\alpha}
(E^{\alpha}_{IV}$) & $S^{\alpha}_{opt}$ &  $S^{\alpha}_{th}$ & $U^{\alpha}_M$ &
$R^{\alpha}_{opt}$ & $\Delta R^{\alpha}_{th}$ & $E^{\alpha}_{PES}$ \\ \hline
(a) Mn$^{4+}$  &   2.56  & 1.73  & \ 47.41 (51.20)  & -44.85 & -45.68  & -38.3 
&-6.55  & -0.83 & 1.2 \\    
\ \ O$^{-}$    &   4.86  & 2.48  & -13.91        &  18.77 &  16.39  & 22.1 & -
3.33 & -2.38 & 3.5 \\     
\ \ \ La$^{4+}$  &  18.36  & 17.63 & \ 49.45 (49.45) & -31.09 & -31.82  & -27.4 
&
-3.68 & -0.73 & 17.0 \\ \hline
(c) Mn$^{4+}$ & 2.62 & 1.95 & 46.83 (51.20) & -44.27 & -44.94 & -38.1 & -6.17 &
-0.67 & 1.2 \\
\ \ O$^-$ & 4.92 & 2.52 & -13.82 & 18.74 & 16.34 & 22.0 & -3.26 & -2.40 & 3.5 \\
\ \ \ La$^{4+}$ & 18.42 & 17.84 & 49.45 (49.45) & -31.03 & -31.61 & -27.4 & -
3.63 &
-0.58 & 17.0 \\    
\end{tabular}
\end{ruledtabular}
\end{table}

\begin{table} 
\caption
{Parameters of the imaginary part of the dielectric function
$\epsilon_2$ \cite{Arima and Tokura,Private} represented by the sum
of Lorentzian shaped bands.}  
\begin{ruledtabular}
\begin{tabular}{cccc}     
$E_i$ & $ \gamma_i$ & ${\nu^2}_{{\bf \rm p}i} $ & $f_i$  \\     
(eV) & (eV) & (eV$^2$) &  \\   \hline 
 1.93 & 1.46 & 1.895 & 0.51\\  
 4.75 & 2.0 & 4.22 & 0.187 \\  
 9.07 & 5.1 & 12.75 & 0.155 \\
\end{tabular}  
\end{ruledtabular}
\end{table}  

\begin{table}  
\caption
{Calculated optical, $E_{opt}$, and thermal, $E_{th}$, energies for the main 
charge-transfer transitions in LaMnO$_3$. 
$S_{opt}$ and $S_{th}$ are
resultant calculated values of the sum of the defect energies 
corresponding the charge-transfer process considered.}  
\begin{ruledtabular}
\begin{tabular}{lclclcc}    
\hspace{1.5cm} CT transition & &Optical energy & Exp. & Thermal energy, & $S_{opt}$ & $S_{th},$ \\  
   & & $E_{opt}$, (eV)  & (eV) & $E_{th}$, (eV) & (eV) & (eV) \\   \hline  
1. Mn$^{4+}$ + Mn$^{3+}$ $\Rightarrow$ Mn$^{3+}$ + Mn$^{4+}$  & &1.33 & - & \ 0.00 &
 -44.35 & \ \ -45.68\\  
2. O$^{-}$ + Mn$^{3+}$ $\Rightarrow$ O$^{2-}$ + Mn$^{4+}$  & &  1.43 (2.29) & 1.93 & 
-0.75 ( 0.12) & -43.50 & \ \ -45.66 \\  
3. Mn$^{4+}$ + O$^{2-}$ $\Rightarrow$ Mn$^{3+}$ + O$^{-}$ & & 2.98 (2.12) & 1.93 & \ 
0.75 (-0.12) & 18.62
&  \ \ \ 16.39 \\     
4. 2Mn$^{3+}$ $\Rightarrow$ Mn$^{4+}$ + Mn$^{2+}$ & Mn(3d) gap & 3.72 & 3.5\cite{Lee}, 3.2 \cite{Takenaka}  &  \
2.68 & -13.06  & \ \ -14.10 \\           
5. Mn$^{3+}$ + O$^{2-}$ $\Rightarrow$ Mn$^{2+}$ + O$^{-}$ & O(2p) - Mn(3d) & 5.61 (4.75) &
4.75 & \ 3.50 & \ 50.15 & \ \ \ 48.04  \\  
6. La$^{3+}$ + O$^{2-}$ $\Rightarrow$ La$^{2+}$ + O$^-$ & O(2p) - La(5d) & 8.93 & 9.07 &  \ 6.47 & \ 42.02 
& \ \ \ 39.56 \\
\end{tabular}
\end{ruledtabular}
\end{table} 

\newpage

\begin{figure}
\includegraphics{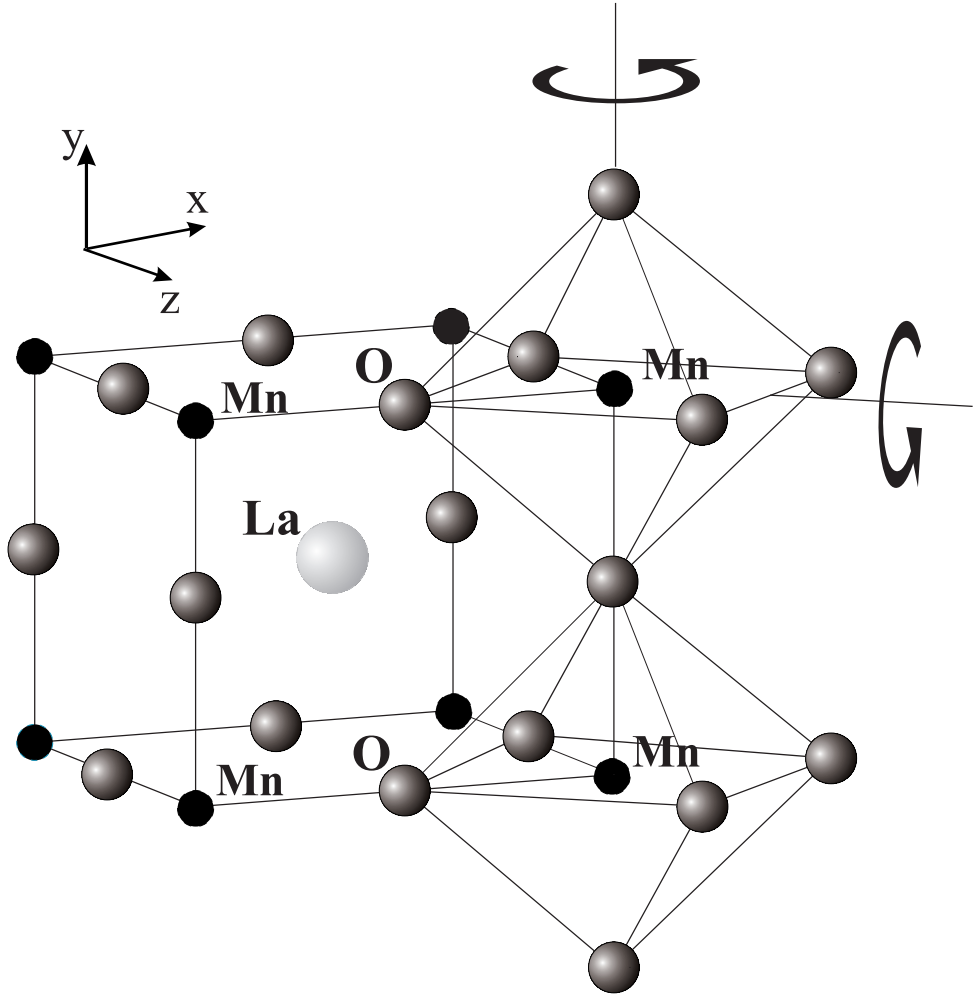}
\caption{The idealized cubic perovskite structure ({\it Pm3m}) of 
LaMnO$_3$ crystal. The orthorhombic {\it Pnma} structure can be obtained 
by two consequent rotations of the MnO$_6$ octahedra around the [010] and [101] 
directions.} 
\label{Figure 1}
\end{figure}
 
\newpage

\begin{figure}
\includegraphics{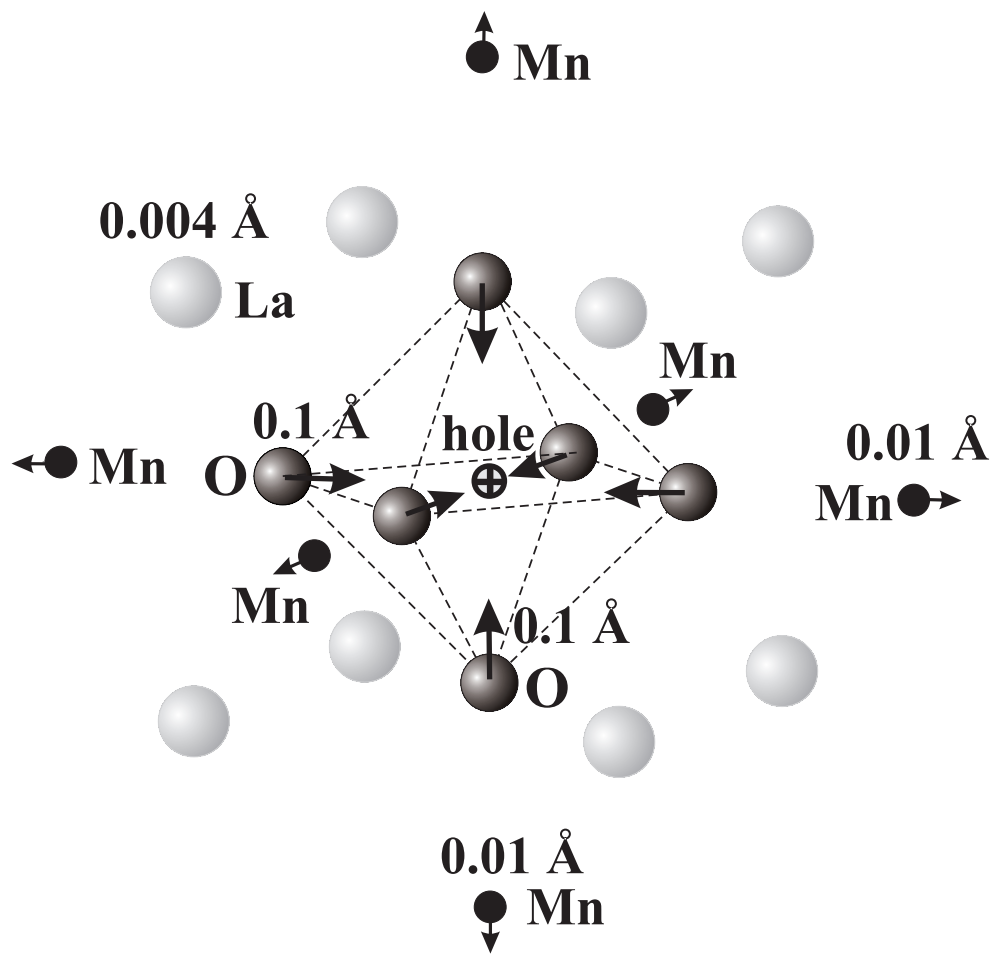}
\caption{The core displacements ($ \geq $ 0.004 \AA) of the ions surrounding 
Mn$^{4+}$ electronic hole defect after full relaxation of cores and shells in 
the LaMnO$_3$ lattice.}
\label{Figure 2}
\end{figure}

\newpage

\begin{figure} 
\includegraphics{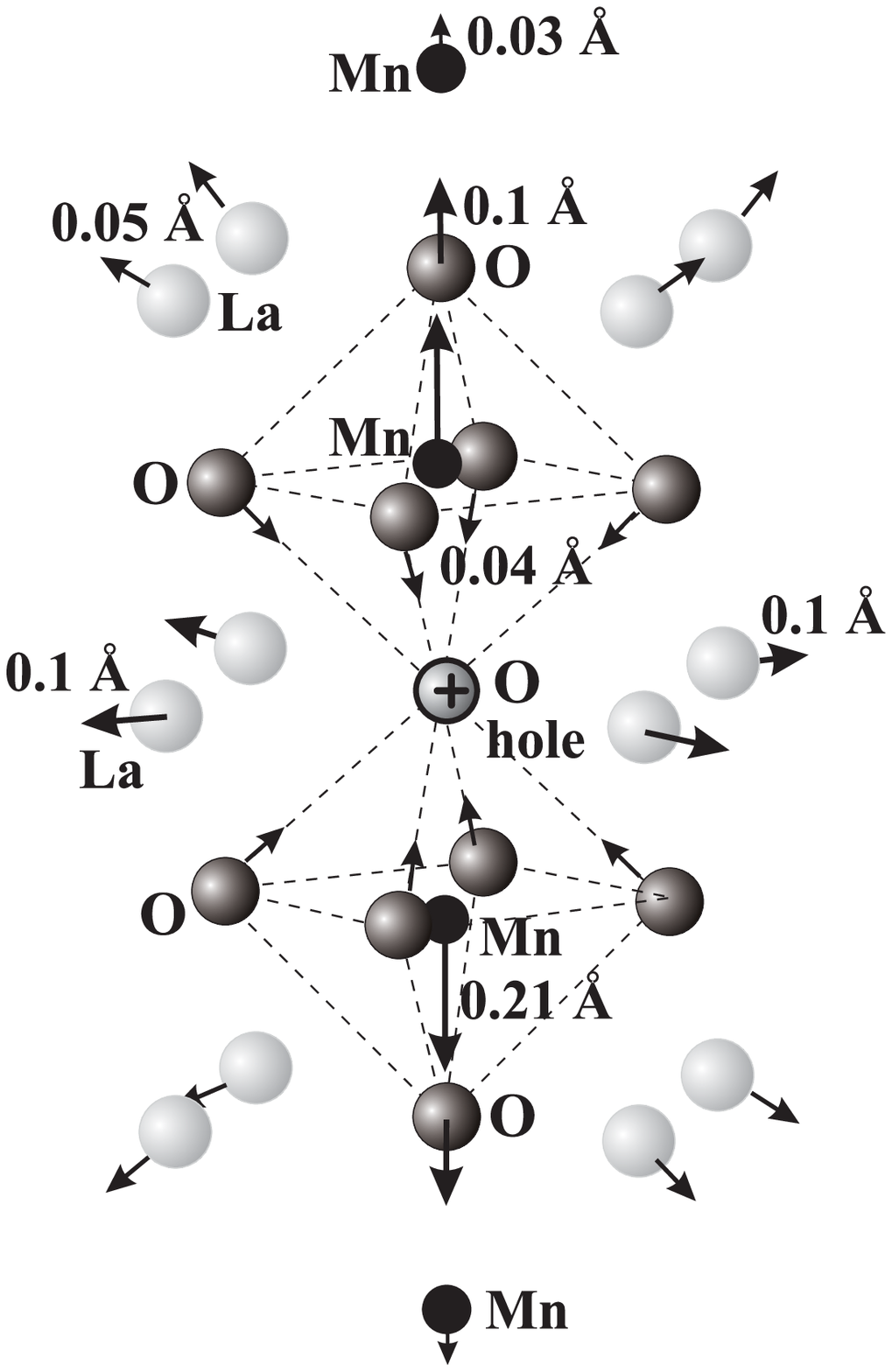}
\caption{The core displacements ($\geq$ 0.03 \AA) of the ions surrounding  
O$^-$ electronic hole defect after full relaxation of cores and shells in the 
LaMnO$_3$ lattice.}
\label{Figure 3}
\end{figure}

\newpage

\begin{figure} 
\includegraphics{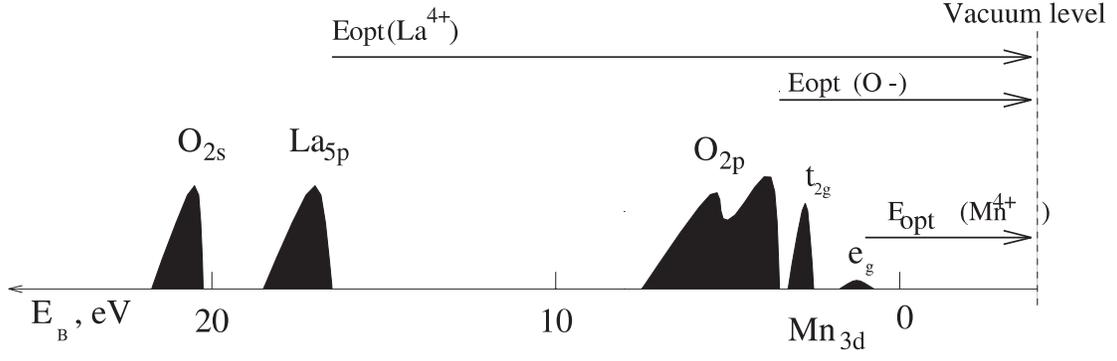}
\caption{A schematic representation of the valence band structure of 
LaMnO$_3$ crystal, showing binding energies \cite{Saitoh,Park} with respect to 
the crystal Fermi level E$_F$. The electron optical excitation processes 
to the vacuum level from the 
Mn(3d) e$_g$, O(2p) and La(5p) valence bands are shown by arrows. 
These optical excitation energies can be compared with experimental 
photoelectron spectroscopy data \cite{Saitoh,Park} and with the calculated 
values of optical energies, $E_{opt}$, for Mn$^{4+}$, O$^-$ and La$^{4+}$ hole 
formation (Table III).}
\label{Figure 4}
\end{figure}

\newpage

\begin{figure}
\includegraphics{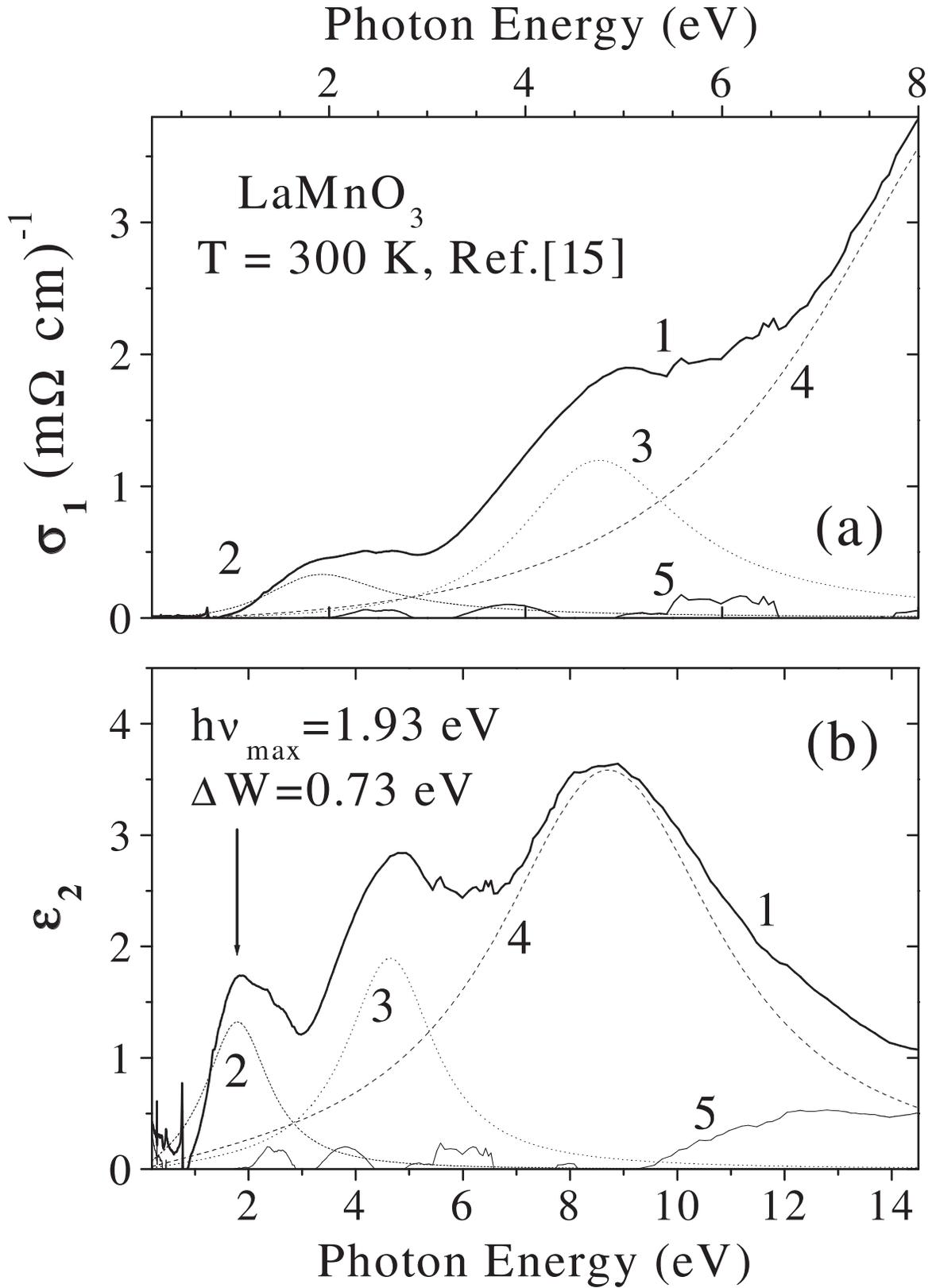}
\caption{(a) - The experimental optical conductivity spectrum of LaMnO$_3$ crystal 
(solid curve 1, T = 300 K) represented by 
the contributions from the three Lorentz oscillators
in accordance with the dispersion analysis of the 
imaginary part of the dielectric function $\epsilon_2$ (b).
(b) - The experimental $\epsilon_2$ spectrum of LaMnO$_3$ 
(solid curve 1, T = 300 K) represented by a sum of 
three main Lorentzian shaped bands: 1.93, 4.75 and 
9.07 eV (drawn by short-dashed 2, dotted 3, and long-dashed 4 curves, respectively).
The rest of $\epsilon_2$ spectrum after subtraction 
of the Lorentzian bands is shown by light line 5. 
The Lorentzian band parameters are given in the Table IV 
together with the estimated oscillator strengths $f_i$.} 
\label{Figure 5}
\end{figure}

\newpage

\begin{figure}
\includegraphics{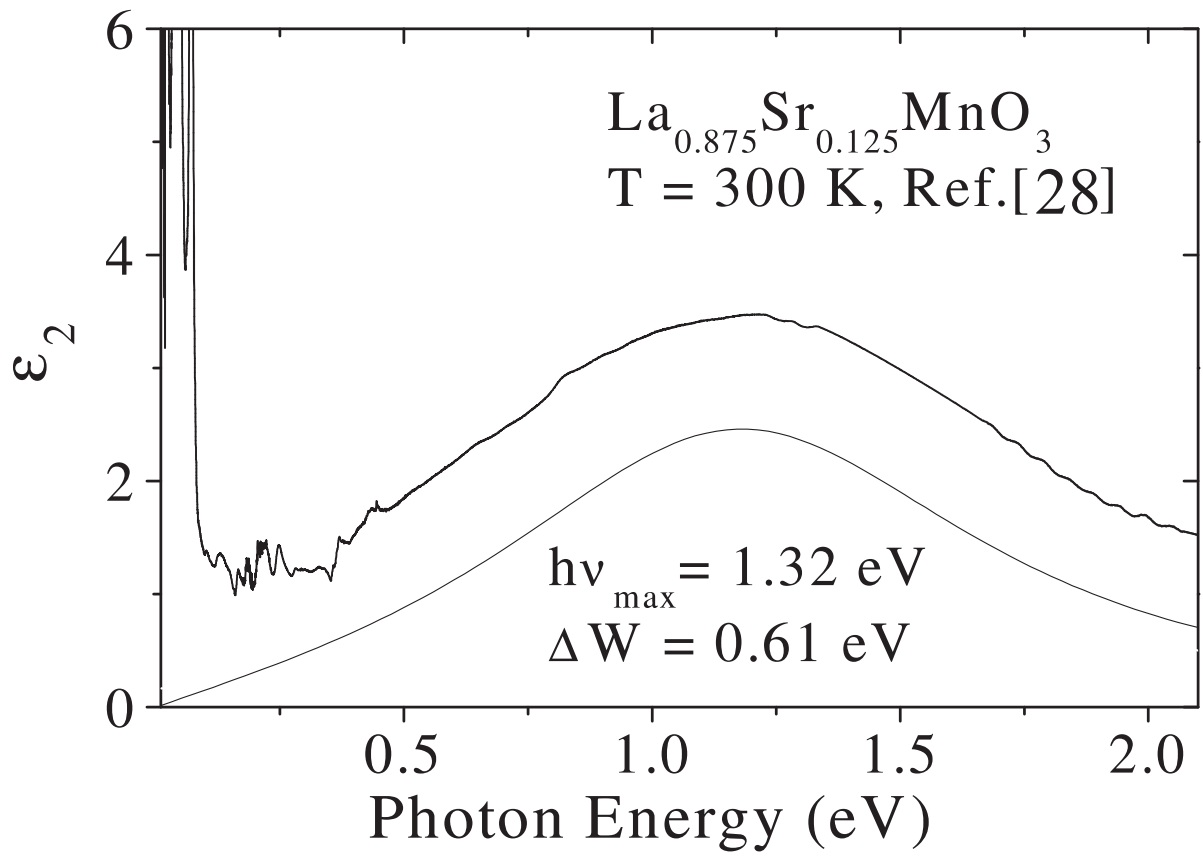}
\caption{The experimental $\epsilon_2$ spectrum of La$_{7/8}$Sr$_{1/8}$MnO$_3$ 
(\cite{Mayr}, T = 300 K) approximated by the Lorentzian shaped band.}
\label{Figure 6}
\end{figure}

\end{document}